\documentstyle[12pt,epsf]{article}

\def \dfrac #1#2 {\displaystyle\frac{#1}{#2}}
\def\be{\begin{eqnarray}}
\def\ee{\end{eqnarray}}
\def\bq{\begin{equation}}
\def\eq{\end{equation}}
\def\ben{\begin{enumerate}}\def\een{\end{enumerate}}

\def\roughly#1{\mathrel{\raise.3ex\hbox{$#1$\kern-.75em
\lower1ex\hbox{$\sim$}}}}

\begin{document}
\def\bra{\langle }
\def\ket{\rangle }
\begin{titlepage}

\vspace{.3cm}
\hfill {\large FTUV-02-0108 ; IFIC-02-10}
\vspace{.2cm}
\begin{center}
\ \\
{\LARGE \bf Generalized Parton Distributions
\\
\vspace{.2cm}
in Constituent Quark Models$\dagger$}
\ \\
\ \\
\vspace{1.0cm}
{
Sergio Scopetta$^{(a)}$ and Vicente Vento$^{(b)}$}
\vskip 0.5cm
{\it (a) Dipartimento di Fisica, Universit\`a degli Studi
di Perugia, via A. Pascoli
06100 Perugia, Italy
\\
and INFN, sezione di Perugia}
\\
{\it (b) Departament de Fisica Te\`orica,
Universitat de Val\`encia, 46100 Burjassot (Val\`encia), Spain
\\
and Institut de F\'{\i}sica Corpuscular,
Consejo Superior de Investigaciones Cient\'{\i}ficas
\\
and 
School of Physics, Korea Institue for Advance Study, Seoul 130-012, Korea}

\end{center}

\vskip 1.0cm
\centerline{\bf Abstract}
\vskip 0.4cm

An approach is proposed to calculate Generalized Parton Distributions
(GPDs) in a Constituent Quark Model (CQM) scenario.
These off-diagonal distributions are obtained 
from momentum space wave functions to be evaluated
in a given non relativistic or relativized CQM. 
The general relations linking the twist-two GPDs
to the form factors and to the leading
twist quark densities are consistently recovered
from our expressions.
Results for the leading twist, unpolarized
GPD, $H$, in a simple harmonic oscillator model, as well
 as in the Isgur and Karl model,
are shown to have the general behavior found in previous 
estimates.
NLO evolution of the obtained distributions, from the low
momentum scale of the model up to the experimental one, is also
shown. Further applications of the proposed formalism 
are addressed.

\vskip 1cm
\leftline{Pacs: 12.39-x, 13.60.Hb, 13.88+e}
\leftline{Keywords:  hadrons, partons, generalized parton distributions, 
quark models.} 
\vspace{.7cm}

{\tt
\leftline {sergio.scopetta@pg.infn.it}
\leftline{vicente.vento@uv.es}}

\vspace{0.7cm}
\noindent{\small$\dagger$Supported in part by SEUI-BFM2001-0262
and by MIUR through the funds COFIN01}
\end{titlepage}

\section{Introduction}

In recent years, Generalized Parton Distributions 
have become one of the main topics of interest in hadronic physics
(for recent reviews, 
see, e.g.,  \cite{jig,rag,pog,burk1,kroll}). 
GPDs are a natural bridge between exclusive processes,
such as elastic scattering, described in terms of form factors,
and inclusive ones, described in terms of structure functions.
As it happens for the usual Parton Distributions (PDFs), the measurement of
GPDs allows important tests of non-perturbative 
and perturbative aspects of the theory, QCD, 
and of phenomenological models
of hadrons. Besides,
GPDs provide us with a unique way to access several 
crucial features of the structure of the nucleon. In particular,
as pointed out first by Ji \cite{ji1,jiprd},
by measuring GPDs
a test of the Angular Momentum Sum Rule
of the proton \cite{jaffe} could be
achieved for the first time, determining
the quark orbital angular momentum contribution to the proton spin.
Therefore,
relevant experimental efforts to measure GPDs,
by means of exclusive electron Deep Inelastic Scattering
(DIS) off the proton, are likely to
take place in the next few years \cite{ceb,esop}.
Besides, GPDs measurements
will be done soon, also through fitting to the available data of the H1 and
ZEUS Collaborations, with the HERMES results used as cross 
checks but excluded from
the fit.            
In this scenario, it becomes urgent to produce 
theoretical predictions for the behavior of these quantities.
Several calculations have been already performed by using different 
descriptions 
of hadron structure: bag models \cite{meln,vv}, soliton models
\cite{pog,goe1}, light-front approaches \cite{mill} and
phenomenological estimates 
based on parametrizations of PDFs
\cite{freund,rad1}.
Besides, an impressive effort has been devoted to study
the perturbative QCD
evolution \cite{scha2,evo} of GPDs, and the GPDs at twist three
accuracy\cite{scha1}.

So far, to our knowledge, no calculations have been performed
in a Constituent Quark Model (CQM), although
a step towards this can be found in \cite{burk1,burk2},
where the non relativistic limit is shortly discussed.
The CQM has a long story of successful
predictions in low energy studies of the electromagnetic
structure of the nucleon.
In the high energy sector,
in order to compare model predictions
with data taken in DIS experiments, 
one has to evolve, according to perturbative QCD, the leading twist
component of the physical structure functions obtained
at the low momentum scale associated with the model.
Such a procedure, already addressed in \cite{pape,jaro}, 
has proven
successful in describing the gross features of 
standard PDFs 
by using different CQM
(see, e.g., \cite{trvv}).
Similar expectations motivated the present study of GPDs.
In this paper, a simple formalism is proposed to calculate
the quark contribution to 
GPDs from any non relativistic or relativized model.
By using such a procedure, the GPDs can be
easily estimated, providing us with an important
tool for the planning 
of future experiments.

The paper is structured as follows.
After the definition of the main quantities
of interest, the proposed calculation scheme
is introduced in the third section. 
Then, results obtained in a simple harmonic oscillator 
model and in the Isgur and Karl model \cite{ik}
are shown in the following section.
NLO perturbative evolution
from the scale of the model to the experimental one
has been evaluated, and results are presented in the fifth section.
Further applications of the procedure, using
relativistic models
and including antiquarks degrees of freedom,
are in progress and will be presented elsewhere \cite{tbp}.
Conclusions are drawn in the last section.

\section{General formalism}

We adopt the formalism introduced
by Ji, who called GPDs 
``Off-forward parton distributions''
in \cite{ji1,jiprd}. 
The connection of these quantities with the  
``Non diagonal'' ones introduced in \cite{rag} is discussed
in \cite{kroll,rad1} and can be easily obtained. 

We are interested in diffractive DIS processes.
The absorption of a high-energy
virtual photon by a quark in a 
hadron target is followed by
the emission of a particle to be later
detected; finally, the interacting quark
is reabsorbed back into the recoiling hadron.
If the emitted and detected particle is, for example, a real photon,
the so called Deeply Virtual Compton Scattering \cite{ji1,jiprd}
process takes place.
Let us think now to a nucleon target, with initial
and final momenta $P$ and $P'$, respectively. 
The GPD $H(x,\xi,\Delta^2)$,
the main quantity
we deal with in the present paper,
is introduced by defining the twist-two part
of the light-cone correlation
function
\begin{eqnarray}
\label{eq1}
\int {d \lambda \over 2 \pi} e^{i \lambda x}
\bra P' | \bar \psi (- {\lambda n \over 2})
\gamma^\mu \psi ({\lambda n \over 2}) | P \ket & = & 
\nonumber
\\
= H(x,\xi,\Delta^2) \bar U(P') \gamma^\mu U(P) & + &
E(x,\xi,\Delta^2) \bar U(P') 
{i \sigma^{\mu \nu} \Delta_\nu \over 2M} U(P) + ...
\end{eqnarray}
\\
where 
$\Delta=P^\prime -P$
is the momentum transfer to the nucleon,
ellipses denote higher-twist contributions,
$\psi$ is a quark field and M is the nucleon mass.
In obtaining Eq. (1), a system of coordinates has been chosen where
the photon 4-momentum, $q^\mu=(q_0,\vec q)$, and $\bar P=(P+P')^\mu/2$ 
are collinear along $z$.
The $\xi$ variable, the so called ``skewedness'', parameterizing
the asymmetry of the process, is defined
by the relation $\xi = - n \cdot \Delta$,
where $n=(1,0,0,-1)/
(2 \Lambda)$ and $\Lambda$ depends on the reference
frame, being, for example, $\Lambda=M/2$ in the nucleon rest frame.
The $\xi$ variable is bounded by 0 and 
$\sqrt{-\Delta^2}/\sqrt{M^2-\Delta^2/4}$. 
Besides, one has $t=\Delta^2=
\Delta_0^2 - \vec{\Delta}^2$. 

In the r.h.s of Eq. (1), the dependence 
of the light-cone correlation
function
on the 
GPDs $H(x,\xi,\Delta^2)$ and $E(x,\xi,\Delta^2)$ is explicitly shown. 
By replacing, in the above equation, $\gamma^\mu$
with the proper Dirac operator,
similar expressions can be derived for defining polarized or chiral
odd GPDs \cite{ji1}.
In the following we will only discuss the unpolarized, chiral even,
twist-two GPD $H(x,\xi,\Delta^2)$.
\\
As explained in \cite{ji1,jiprd},
unlike the usual PDFs,
which have the physical meaning of a momentum density in the
Infinite Momentum Frame (IMF), GPDs have the meaning of
a probability amplitude. 
They describe the amplitude for finding a quark with momentum fraction
$~~x+\xi/2$ (in the IMF) in a nucleon 
with momentum $(1+\xi/2) \bar P$
and replacing it back into
the nucleon with a momentum transfer $\Delta^\mu$.
Besides, when the quark longitudinal momentum fraction 
$x$ of the average nucleon momentum $\bar P$
is less than $-\xi/2$, GPDs describe antiquarks;
when it is larger than $\xi/2$, they describe quarks;
when it is between $-\xi/2$ and $\xi/2$, they describe 
$q \bar q$ pairs.\footnote{Note that in going from Ref.
\cite{jiprd} to Ref. \cite{jig} Ji has redefined $\xi/2$
by $\xi$.}

The region $|x| \geq \xi/2$ is often called the DGLAP region,
since the $Q^2$ evolution of GPDs is governed there
by the DGLAP equations of perturbative QCD \cite{dglap};
the region $|x| \leq \xi/2$ is called the ERBL region,
because there the $Q^2$ evolution is ERBL-like \cite{erbl}.
One should keep in mind that, besides the variables
$x,\xi$ and $\Delta^2$ explicitely shown, GPDs depend,
as the standard PDFs, on the momentum scale $Q^2$ at which they are
measured or calculated. For an easy presentation,
such a dependence will be 
omitted in the next two sections of the paper, while it will be
properly discussed in the last one. 

There are two natural limits for $H(x,\xi,\Delta^2)$ : 
i) when $P^\prime=P$, i.e., $\Delta^2=\xi=0$, the so called
``forward'' or ``diagonal'' limit, one recovers
the usual PDFs
\begin{eqnarray}
H(x,0,0)=q(x)~;
\end{eqnarray}
ii)
the integration over $x$ is independent on $\xi$
and yields the Dirac 
Form Factor (FF)
\begin{eqnarray}
\int dx H(x,\xi,\Delta^2) = F_1(\Delta^2)~. 
\end{eqnarray}
Any model estimate of the GPDs has to respect these two 
crucial constraints.

\section{A non relativistic scheme}

Our aim now is to evaluate
the Impulse Approximation (IA) expression
for 
$H(x,\xi,\Delta^2)$, suitable to perform CQM
calculations.

Let us start from Eq. (1).
Substituting the quark fields in the left-hand-side one has 

\begin{eqnarray}
\int {d \lambda \over 2 \pi} e^{i \lambda x}
\bra P' | \bar \psi (- {\lambda n \over 2})
\gamma^\mu \psi ({\lambda n \over 2}) | P \ket & = & 
\nonumber
\\
\int {d \lambda \over 2 \pi} e^{i \lambda x}
\bra P' |
\sum_{r,r',\vec k} [ 
d_{r'}(\vec k + \vec \Delta) \bar v_{r'}(\vec k + \vec \Delta)
e^{i (k+\Delta)\cdot {\lambda n \over 2} }
& + &
b^+_{r'}(\vec k + \vec \Delta) \bar u_{r'}(\vec k + \vec \Delta)
e^{-i (k+\Delta) \cdot {\lambda n \over 2} }
] 
\nonumber
\\
\gamma^\mu [
b_{r}(\vec k) u_{r}(\vec k)
e^{-i k \cdot {\lambda n \over 2} }
& + &
d^+_{r}(\vec k )  v_{r}(\vec k)
e^{i k \cdot {\lambda n \over 2} }] | P \ket 
\nonumber
\end{eqnarray}
and, taking into account the quarks degrees of freedom only,
it becomes

\begin{eqnarray}
C = \sum_{r,r',\vec k} 
\int {d \lambda \over 2 \pi} e^{i \lambda x}
\bra P' |
\bar u_{r'}(\vec k + \vec \Delta) b^+_{r'}(\vec k + \vec \Delta)
e^{-i (k+\Delta) \cdot {\lambda n \over 2} }
\gamma^\mu  
b_{r}(\vec k) u_{r}(\vec k)
e^{-i k \cdot {\lambda n \over 2} }
| P \ket~. 
\nonumber
\end{eqnarray}
Using IA and integrating
over $\lambda$ yields

\begin{eqnarray}
C & \simeq & \sum_i \sum_{r,r',\vec k} 
\int {d \lambda \over 2 \pi} e^{i \lambda (x
- { n \over 2} \cdot (2k + \Delta))}
\bra P' |
\bar u_{r'}(\vec k + \vec \Delta)
b^+_{i,r'}(\vec k + \vec \Delta)
\gamma^\mu
b_{i,r}(\vec k)
u_{r}(\vec k)
| P \ket =
\nonumber
\\
& = & \sum_i \sum_{r,r',\vec k} 
\delta \left (x - { n \over 2} \cdot (2k + \Delta) \right )
\bra P' |
\bar u_{r'}(\vec k + \vec \Delta)
b^+_{i,r'}(\vec k + \vec \Delta)
\gamma^\mu
b_{i,r}(\vec k)
u_{r}(\vec k)
| P \ket~. 
\nonumber
\end{eqnarray}
Let us introduce $\xi = - n \cdot \Delta$,
so that  Eq. (1) reads now:

\begin{eqnarray}
& & \sum_i \sum_{r,r',\vec k}  
\delta \left (x + { \xi \over 2} - n \cdot k \right )
\bra P' | 
\bar u_{r'}(\vec k + \vec \Delta)
b^+_{i,r'}(\vec k + \vec \Delta)
\gamma^\mu 
b_{i,r}(\vec k)
u_{r}(\vec k)
| P \ket =
\nonumber
\\
& = &
H(x,\xi,\Delta^2) \bar U(P') \gamma^\mu U(P)  + 
E(x,\xi,\Delta^2) \bar U(P') 
{i \sigma^{\mu \nu} \Delta_\nu \over 2M} U(P)~,
\label{iahe}
\end{eqnarray}
which holds exactly if the antiquark degrees of freedom
are not considered. In fact, the l.h.s. is evaluated in IA
and the r.h.s. is the leading-twist
part of the light-cone correlation function, 
so that they have the same physical content.

%
%
%
By taking the zero-components in the left and right
hand sides of Eq. (\ref{iahe}), considering
a process with ${\vec \Delta}^2 \ll M^2$,
one immediately sees that the contribution of the term proportional
to $E(x,\xi,\Delta^2)$, in the right hand side of 
Eq. (\ref{iahe}), becomes negligibly small, so that
$H(x,\xi,\Delta^2)$ is given by:
\begin{eqnarray}
H(x,\xi,\Delta^2)  =  
\sum_i \sum_{r',r,\vec k}  
\delta \left ( x + { \xi \over 2} - {k^+ \over M} \right )
\bra P' | u_{r'}^+(\vec k + \vec \Delta)
b^+_{i,r'}(\vec k + \vec \Delta)
b_{i,r}(\vec k) u_{r}(\vec k)
| P \ket~,
\label{hket}
\end{eqnarray}
where $k^+ = k_0 + k_3$ has been introduced.
In order to evaluate
this expression by means of a 
CQM, one
has to relate it to
nucleon wave functions.
In a non relativistic framework, if
the normalization of
the nucleon states is chosen to be
\begin{eqnarray}
\bra P' | P \ket = (2 \pi)^3 \delta( \vec P' - \vec P )~,
\nonumber
\end{eqnarray}
for a symmetric wave function
(as is the case in a quark model
once color has been taken into account), one has
(see, e.g. , \cite{muld})
\begin{eqnarray}
\sum_i \sum_{r',r} \bra P' |
u_{r'}^+(\vec k + \vec \Delta)
b^+_{i,r'}(\vec k + \vec \Delta)
b_{i,r}(\vec k) 
u_r(\vec k)
| P \ket  
& = & 
3 \int \psi^*(\vec k_1, \vec k_2, \vec k + \Delta)
\psi(\vec k_1, \vec k_2, \vec k) d \vec k_1 d \vec k_2 =
\nonumber
\\
& = & \int e^{i ((\vec k + \vec \Delta) \vec r
-\vec k \vec r' )} \rho(\vec r, \vec r') d \vec r
d \vec r'~,
\nonumber
\\
& = & \tilde n(\vec k , \vec k + \vec \Delta) 
\label{nkk}
\end{eqnarray}
where the one-body non diagonal charge density
\begin{eqnarray}
\rho(\vec r, \vec r') 
=  
\int \psi^*(\vec r_1, \vec r_2, \vec r') \,
\psi(\vec r_1, \vec r_2, \vec r) \, d \vec r_1 \, d \vec r_2 
\label{rho}
\end{eqnarray}
and the one-body non-diagonal momentum distribution
$\tilde n(\vec k , \vec k + \vec \Delta)$ have been introduced.
In terms of the latter quantity, Eq. (\ref{hket}) can be written
\begin{eqnarray}
H(x,\xi,\Delta^2)  = \int d \vec k\,\,\,
\delta \left (x + { \xi \over 2} - {k^+ \over M}  \right )\,
\tilde n(\vec k , \vec k + \vec \Delta)~. 
\label{hnk}
\end{eqnarray}

The above equation, which is our basic result,
allows the calculation of 
$H(x,\xi,\Delta^2)$ in any CQM, and
it naturally verifies the two crucial constraints, Eqs. (2) and (3). 
In fact,
the unpolarized quark density, $q(x)$, in the IA
is recovered in the forward limit
when $\Delta^2=\xi=0$:
\begin{eqnarray}
q(x) = H(x,0,0)=\int d \vec k \, n(\vec k) \, \delta \left ( x - {k^+ \over M}
\right )
\label{hf}
\end{eqnarray}
so that the constraint Eq. (2) is fulfilled.
In the above equation, $n(\vec k)$ is the momentum distribution
of the quarks in the nucleon:
\begin{eqnarray}
n(\vec k) =  \int e^{i \vec k \cdot ( \vec r - \vec r')}
\rho(\vec r, \vec r') d \vec r
d \vec r'~.
\end{eqnarray}

As is well known,
the relation between the quark momentum distribution
and the quark unpolarized density, Eq. (\ref{hf}), can be found by analyzing,
in IA,
the handbag diagram, i.e., the leading twist part
of the full DIS process
(see, e.g., \cite{muld,traini}),
assuming that the interacting quark is on-shell.
So, from Eq. (\ref{hnk}), derived as the non relativistic reduction
of the light-cone correlation function in the IA, 
the quark density appears as the forward limit.
Besides, integrating Eq. (\ref{hnk}) over $x$, 
one trivially
obtains
\begin{eqnarray}
\int d x H(x,\xi, \Delta^2) =
\int d \vec r e^{i \vec \Delta \vec r} \rho(\vec r)
~,
\nonumber
\end{eqnarray}
where $\rho(\vec r) = \lim_{\vec r' \rightarrow \vec r}
\rho(\vec r', \vec r)$ is the quark charge density.
The r.h.s. of the above 
equation gives the IA definition of the charge FF 
\begin{eqnarray}
\int d \vec r e^{i \vec \Delta \vec r} \rho(\vec r)=
F(\Delta^2)
~,
\label{ffr}
\end{eqnarray}
so that, recalling that
$F(\Delta^2)$ coincides
with
the non relativistic limit of the Dirac FF $F_1(\Delta^2)$,
the constraint
Eq. (3) is immediately fulfilled. 

Let us introduce now the following sets of
conjugated intrinsic coordinates
\begin{eqnarray}
\vec R = { 1 \over \sqrt{3} } ( \vec{r_1} + \vec{r_2} + \vec{r_3} )
& \leftrightarrow &  
\vec K = { 1 \over \sqrt 3} ( \vec{k_1} + \vec{k_2} + \vec{k_3} )
\nonumber
\\
\vec \rho = { 1 \over \sqrt 2} ( \vec{r_1}  - \vec{r_2} )
& \leftrightarrow &  
\vec{k_{\rho}} = { 1 \over \sqrt 2} ( \vec{k_1} -  \vec{k_2} )
\nonumber
\\
\vec \lambda = { 1 \over \sqrt 6} ( \vec{r_1} + \vec{r_2} - 2 \vec{r_3} )
& \leftrightarrow &  
\vec{k_{\lambda}} = { 1 \over \sqrt 6} ( \vec{k_1} + \vec{k_2} - 2 \vec{k_3} )
\nonumber
\end{eqnarray}
in terms of which, in a coordinate system
where $\vec R=0$,
the FF Eq. (\ref{ffr}) can be written \cite{burk2,mmg}
\begin{eqnarray}
\label{ffwf}
F(\Delta^2) = 
\int d \vec{k_{\rho}}  \, d \vec{k_{\lambda}} \,
\psi^* \left( \vec{k_{\rho}}, \vec{k_{\lambda}} - { \sqrt {2 /
3} } \, \vec \Delta \right ) 
\psi \left ( \vec{k_{\rho}}, \vec{k_{\lambda}} \right )
\end{eqnarray} 
and 
$H(x,\xi,\Delta^2)$, Eq. (\ref{hnk}),
by substituting Eq. (\ref{rho}) into Eq. (\ref{nkk}), reads
\begin{eqnarray}
\label{jac}
H(x,\xi,\Delta^2) = \int d \vec{k_{\rho}} \,  d \vec{k_{\lambda}} 
\, \psi^*\left(\vec{k_{\rho}}, \vec{k_{\lambda}} - { \sqrt {2 /
3} }\, \vec \Delta \right) 
\psi (\vec{k_{\rho}}, \vec{k_{\lambda}} )
\, \delta \left( x + {\xi \over 2} - 
{k_3^+ \over M} \right )~.
\end{eqnarray}
One immediately realizes that Eq. (\ref{ffwf}) is obtained from 
Eq. (\ref{jac})
by performing the $x$ integration.

With respect to Eq. (\ref{hnk}),
a few caveats are necessary.
First of all, 
due to the use of CQM wave functions, which contain only
constituent quarks (and also antiquarks in the case of mesons),
only the quark (and antiquark) contribution to the
GPDs can be evaluated, i.e.,
only the region $x \geq \xi/2$ 
(and also $x \leq -\xi/2$ for mesons) can be explored.
In order to introduce the study of the sea region
($ - \xi / 2 \leq x \leq \xi/2$), 
the model has to be enriched.
To this respect, calculations including
a substructure
of the constituent quark, as proposed by several authors
\cite{acmp,gianni,scopetta},
are in progress and will be presented elsewhere \cite{tbp}.

Secondly, we remind that
Eq. (\ref{hnk}) holds 
under the condition
$\Delta^2 \ll M^2$.
If one wants to treat more general processes,
such a condition can be easily relaxed by keeping the terms
of $O(\Delta^2/M^2)$ in going from Eq. (\ref{iahe}) to 
Eq. (\ref{hket}). At the same time, an expression to 
evaluate $E(x,\xi,\Delta^2)$ could be readily
obtained. 

Finally,
in the argument of the $\delta$ function
in Eq. (10), due to the approximations used, the $x$ variable
is not defined in its natural support, i.e. it can be
larger than 1 and smaller than $\xi/2$.
Several prescriptions have been proposed in the past
to overcome such a difficulty in the standard PDFs case \cite{jaro,trvv}.
The support violation is small for the calculations that will be shown here. 
However one has to be cautious in interpreting the results after pQCD evolution
is perfomed. A deeper discussion of this issue is beyond the
scope of the present work \cite{tbp}.

We stress that our definition of  $H(x,\xi,\Delta^2)$
in terms of CQM
momentum space wave functions can be easily generalized
to other GPDs, and the relation
of the latter quantities with other FFs (for example the magnetic
one) and
other PDFs (for example the polarized quark density) \cite{tbp}
can be recovered.
Therefore the proposed scheme allows one to calculate
GPDs by using any non relativistic or relativized \cite{gianni} CQM, and it is also
suitable to be implemented by corrections due to a
possible finite size and complex structure of the
constituent quarks, as proposed by several authors
\cite{acmp,gianni,scopetta}. 

\section{Results in non relativistic quark models}

As an illustration,
in this section we present the results of our approach
in the CQM of Isgur and Karl (IK) \cite{ik}.
In this model
the proton wave function is obtained in a 
one gluon exchange potential
added to a confining harmonic oscillator (h.o.) one;
including contributions up to the $2 \hbar \omega$ shell, 
the proton state is given by the
following admixture of states
\begin{eqnarray}
|N \rangle = 
a_{\cal S} | ^2 S_{1/2} \rangle_S +
a_{\cal S'} | ^2 S'_{1/2} \rangle_{S} +
a_{\cal M} | ^2 S_{1/2} \rangle_M +
a_{\cal D} | ^4 D_{1/2} \rangle_M~,
\label{ikwf}
\end{eqnarray}
where the spectroscopic notation $|^{2S+1}X_J \rangle_t$, 
with $t=A,M,S$ being the symmetry type, has been used.
The coefficients were determined by spectroscopic properties to be
\cite{mmg}: 
$a_{\cal S} = 0.931$, 
$a_{\cal S'} = -0.274$,
$a_{\cal M} = -0.233$, $a_{\cal D} = -0.067$.
 
If $a_{\cal S} = 1$ and 
$a_{\cal S'} = 
a_{\cal M} = a_{\cal D} = 0$, the simple h.o. model
is recovered.
Let us now calculate the GPD $H$ in the IK model
by using Eq. (\ref{jac}).  
The different components appearing in the momentum
space wave functions, obtained from Eq. (\ref{ikwf})
in the IK model, can be found in 
\cite{mmg,ik2};
for the h.o. model,
the corresponding wave function
in momentum space reduces to \cite{traini,mmg}
\begin{eqnarray}
\label{howf}
\psi (\vec{k_{\rho}}, \vec{k_{\lambda}} ) 
= { e^{- { { k_{\rho}^2 + k_{\lambda}^2} \over 2 \alpha^2 } }  
\over \pi^{3/2} \alpha^3 }~,
\end{eqnarray}
where the h.o. parameter can be fixed
to $\alpha^2=1.35 f^{-2}$
in order to reproduce the low $t$ behavior of the charge
FF, i.e., the r.m.s. value of the proton radius.
\\
The results in the IK model for the 
GPD $H(x,\xi,\Delta^2)$, for the flavours $u$ and $d$,
respectively, neglecting in (\ref{ikwf})
the small $D$-wave contribution, are found to be:

\begin{eqnarray}
H_u(x,\xi,\Delta^2) 
& = & 3 { M \over \alpha^3}  
\left( {3 \over 
2 \pi } \right)^{3/2} 
e^{ - { \Delta^2 \over 3 \alpha^2 }} 
\int dk_x \int dk_y \,
f_0(k_x,k_y,x,\xi,\Delta^2) 
\nonumber
\\
& \times &
\left ( f_s(k_x,k_y,x,\xi,\Delta^2)
+ {\tilde f} (k_x,k_y,x,\xi,\Delta^2) \right )
\label{iku}
\end{eqnarray}

\begin{eqnarray}
H_d(x,\xi,\Delta^2) 
& = & 3 { M \over \alpha^3}  
\left( {3 \over 
2 \pi } \right)^{3/2} 
e^{ - { \Delta^2 \over 3 \alpha^2 }} 
\int dk_x \int dk_y \,
f_0(k_x,k_y,x,\xi,\Delta^2) 
\nonumber
\\
& \times &
\left ( {1 \over 2} f_s(k_x,k_y,x,\xi,\Delta^2)
- {\tilde f} (k_x,k_y,x,\xi,\Delta^2)\right )
\label{ikd}
\end{eqnarray}

with

\begin{eqnarray}
f_0(k_x,k_y,x,\xi,\Delta^2) =
{ \bar k_0 \over \bar k_0 + \bar k_z }
f_\alpha(\Delta_x,k_x)
f_\alpha(\Delta_y,k_y) f_\alpha(\Delta_z,\bar k_z)~,
\end{eqnarray}
\begin{eqnarray}
f_\alpha(\Delta_i,k_i) = e^{ - {1 \over \alpha^2} \left( {3 \over 2} k_i^2  
+ k_i\Delta_i \right) }~~,
\end{eqnarray}
\begin{eqnarray}
\bar{k_z} = { M^2(x + \xi/2)^2 - ( m^2 + k_x^2 + k_y^2)
\over 2 M ( x + \xi/2) }~,
\end{eqnarray}

\begin{eqnarray}
\label{fs}
f_s(k_x,k_y,x,\xi,\Delta^2) & = & {2 \over 3} a_s^2 + a_{s'}^2
\left[ {5 \over 6} - {k^2 \over \alpha^2} + { 1 \over 2} 
{k^4 \over \alpha^4} + { 2 \over {3 \alpha^2}}
\left ( {\Delta^2 \over 3} + \vec \Delta \cdot \vec k \right )
\left ( {k^2 \over \alpha^2} -1 \right ) \right ] \nonumber
\\  
& + & a_M^2 \left [ {5 \over 12} - {1 \over 2} {k^2 \over \alpha^2}  
+ { 1 \over 4} {k^4 \over \alpha^4} +
{ 2 \over 9} {k \over \ \alpha^2} \sqrt{ { 9 \over 4} k^2 +
\Delta^2 + 3 \vec \Delta \cdot \vec k } \right.
\nonumber
\\
& + & \left. { 1 \over {3 \alpha^2}}
\left ( {\Delta^2 \over 3} + \vec \Delta \cdot \vec k \right )
\left ( {k^2 \over \alpha^2} -1 \right ) \right ]
\\
& + & a_S a_{S'} { 2 \over \sqrt{3}} 
\left [ \left ( 1 - {k^2 \over \alpha^2} \right )
- { 2 \over {3 \alpha^2}}
\left ( {\Delta^2 \over 3} + \vec \Delta \cdot \vec k \right )
\right ] \nonumber
\end{eqnarray}

\begin{eqnarray}
\label{ft}
\tilde f(k_x,k_y,x,\xi,\Delta^2) & = & - a_S a_{S'} { 2 \over \sqrt{3}} 
\left [ \left ( 1 - {k^2 \over \alpha^2} \right )
- { 2 \over {3 \alpha^2}}
\left ( {\Delta^2 \over 3} + \vec \Delta \cdot \vec k \right )
\right ] \nonumber \\
& - & a_M a_{S'} { 1 \over \sqrt{2}} 
\left [ { 1 \over 6} - {k^2 \over \alpha^2} +
{1 \over 2} {k^4 \over \alpha^4}
- { 2 \over {3 \alpha^2}}
\left ( {\Delta^2 \over 3} - \vec \Delta \cdot \vec k \right ) \right.
\nonumber
\\
& + & \left. { 2 k^2 \over {3 \alpha^4}}
\left ( {\Delta^2 \over 3} + \vec \Delta \cdot \vec k \right )
\right ]
\end{eqnarray}

and $\bar k_0 = \sqrt{ m_q^2+k_x^2+k_y^2+\bar k_z^2}$, i.e.,
the interacting quark has been assumed to be on shell.

A few comments are in order:
\begin{itemize}

\item{The $x$-integration of Eqs. (\ref{iku}) and (\ref{ikd}) give the
$u$ and $d$ flavor contribution to the proton charge FF in the IA, 
respectively,
as given in \cite{ik2}};

\item{In the forward limit, $\xi=0$, $\Delta^2=0$, Eqs (\ref{iku})
and (\ref{ikd}) give the distributions $H_u(x,0,0)=u(x)$ 
and $H_d(x,0,0)=d(x)$ in the Isgur
and Karl model, according to the findings of Ref. \cite{tcm}.}

\end{itemize}

The results in the simple h.o. model can be immediately found from the 
ones presented above, just putting 
$a_{\cal S} = 1$, $a_{\cal S'} = a_{\cal M} = 0$ in Eqs. (\ref{fs})
and (\ref{ft}). In particular, by using the wave function Eq. (\ref{howf})
in Eq. (\ref{ffwf}), one gets trivially
\begin{eqnarray}
F(\Delta^2) = e^{- {\Delta^2 \over 6 \alpha^2 }}~.
\end{eqnarray}
Besides, taking the ``forward'' limit, $\Delta^2=\xi=0$, of
Eq. (\ref{jac}) and substituting in the w.f.
Eq. (\ref{howf}), or, which is the same, taking
the forward limit of Eqs. (\ref{fs}) and (\ref{ft}) with
$a_{\cal S} = 1$, $a_{\cal S'} = a_{\cal M} = 0$,
and performing analitically the integrations in Eqs.
(\ref{iku}) and (\ref{ikd})
one easily obtains:
\begin{eqnarray}
H(x,0,0) 
=  {2 \pi M \over \alpha^3}  \left ( {3 \over 2 \pi} \right) ^{3/2} 
\int_{k_-(x)}^\infty 
dk k e^{-{ 3 k^2 \over 2 \alpha^2} }
\end{eqnarray}
where the integration limit  $k_-(x)$ is
\begin{eqnarray}
k_-(x)= { M \over 2} \left [ x - {m_q^2 \over M^2}{1 \over x} \right ]
\end{eqnarray}
and $m_q$ is the quark mass.
This is the same expression obtained 
in \cite{traini}
starting from Eq. (9), using
$n(\vec k)$ corresponding to the present model (called ``model 1'' 
in \cite{traini}):
\begin{equation}
n(\vec k) = \left ( { 3 \over  2\pi} \right )^{3/2}  
{e^{-{3 k^2 \over 2 \alpha^2}} \over \alpha^3}~.
\end{equation}

Results are shown in  Figs. 1 to 5.

The behavior of the proton charge FF in the IK model is shown in Fig. 1.
It is known \cite{mmg} that such a FF
underestimates the data for large values of $\Delta^2$.
 
Results for the GPD $H$ at the low momentum
scale $\mu_o^2$ associated with the model, are shown in Figs. 2 to 5.
A value of $m_q \simeq M/3$ has been used for all the estimates.
Note that all the results are shown in the $x \geq \xi/2$ region.
We reiterate in fact that what has been calculated so far 
is the valence quark contribution to the full GPD $H$, so that we
can provide estimates only in the positive DGLAP region. 

In Fig. 2, the $\Delta^2$ dependence of our results for the
$u$ and $d$ flavors is shown.
The forward, $\Delta^2=\xi=0$ limit corresponds to the full line.
One immediately realizes that a strong $\Delta^2$ dependence is found,
in comparison with other estimates, for example with the one
predicted in \cite{meln}. This has to do with
the already discussed strong $t$ dependence 
of the FF in the IK model.

In Figs. 3 and 4 we have the full $\Delta^2$ and $\xi$ dependences.
These findings, particularly clear from the three-dimensional plots
in Fig. 4,
are similar to the ones obtained in \cite{meln},
although the $\xi$ dependence is a little stronger.

The scale of the left and right panels in Figs. 2 and 3
has been chosen in such a way that they would look exactly
the same if the model were SU(6) symmetric
(in fact, in that case, one would have $H_u = 2 H_d$).
The observed difference clearly shows to what extent 
the $SU(6)$ symmetry is broken in the IK model.

In Fig. 5, the comparison between the predictions of the IK
and of the simple h.o. models is shown. As an example,
results are presented for the $d$ distribution at $\xi=0$. 

\section{QCD evolution of the model calculations}

According to a well estblished \cite{pape,jaro},  
widely used scheme (see, for example, \cite{trvv}), the
results shown so far for $H(x,\xi,\Delta^2)$ correspond to
the low momentum scale $\mu_o^2$ associated with the model, and 
in order to compare them with the data of future experiments, 
one has to evolve them to experimental, high-momentum scales.
We next proceed to do so. 

As already mentioned in the Introduction, an impressive effort has 
been devoted to studies of QCD evolution properties of GPDs,
an essential feature to understand their physical content and
to obtain the correct information from experiments.
The QCD evolution of GPDs is presently known in both the DGLAP
and ERBL regions, up to NLO accuracy 
\cite{scha2,evo}.

The evolution of the results presented in the previous section
has been carried on by using an evolution code kindly provided by
Freund and Mc Dermott, adapted by us to our specific case.
The features and performances of such a code are described
in \cite{evo}, and an interface to it is available
at the web site \cite{web}. In order to be used as input
in the evolution code, our GPDs have been translated  into
the Golec-Biernat, Martin off-diagonal parton distributions
$\hat{\cal F}(x_1,\zeta)$ \cite{g-bm}.
The new distribution and the
variables $x_1$ and $\zeta$ are obtained 
from our quantities $H$, $x$ and $\xi$ according to definitions given
in \cite{g-bm}. Once the evolution has been performed,
results have been translated back to our notation to allow
a consistent presentation.

The scale to be associated with the model is a low one,
and the choice of its value is part of the model assumptions.
We have chosen here $\mu_o^2=0.34 GeV^2$, corresponding
to the initial scale of the so called valence scenario
of \cite{grv}, since our input at the scale of the model
is given by the valence quark contribution only.
A thorough discussion about the choice of the initial scale
can be found in \cite{trvv}. Of course, being the starting scale
so low, it becomes very important to perform the evolution
as accurate as possible. From this point of view, the NLO level, 
used here, allows a safe result. 

The evolved GPDs at $Q^2=10$ GeV$^2$ are shown in Figs. 6 and 7,
for two values of $\xi$ and $\Delta^2$, for the Non Singlet
(NS) part of the $u$ and $d$
flavor distributions, together with the input at the initial scale.
One should notice again that only the positive DGLAP region is shown,
for both the input and the evolved distributions. As happens in the 
conventional PD case, and is easily understood, evolution lowers 
the  mean $x$ of the distribution, i.e., the partons accumulate near 
$x= \xi/2$. 

In Fig. 8 a three-dimensional plot of the $Q^2$ evolution, for
fixed $\xi=0.1$ and fixed $\Delta^2=-0.2$ GeV$^2$, is also shown.
 
\section{Conclusions}

Generalized Parton Distributions (GPDs) are a useful tool
to access several relevant features of the structure
of the nucleon, such as the angular momentum sum rule.
A systematic theoretical study of many aspects
of these objects started few years ago and it is being
carried on in these years. The future experimental 
effort to measure GPDs is also ambitious,
and, in this respect, theoretical estimates
will be necessary for the planning of future
experiments.
In the present paper we propose a  general formalism
to investigate GPDs by means of non relativistic or relativized
Constituent Quark Models. Starting from the general field-theoretical
definition of the related light-cone correlation function,
by performing an Impulse Approximation analysis and
the non relativistic limit, the unpolarized, leading-twist
GPD $H(x,\xi,\Delta^2)$ is obtained in terms
of the nucleon wave functions in momentum space.
From its expression,
the quark momentum density is recovered as the forward limit,
and the charge Form Factor as its $x$ integral.
Results for the valence quark contribution to  $H(x,\xi,\Delta^2)$,
in a simple harmonic oscillator model, as well as
in the Isgur and Karl constituent quark model,
are shown to have the general behavior obtained in previous 
estimates. NLO evolution to high experimental scales
of the low momentum results obtained in the model
has been performed.
A proper treatment of the ERBL region within a constituent
picture is presently under investigation.

The proposed approach 
can have many interesting developments,
such as the calculation of other GPD functions
and DVCS observables, the use of
relativistic models and
the addition of effects due to a 
possible finite size and complex structure of the
constituent quarks, as proposed by several authors.

\section*{Acknowledgements}

Many useful discussions with Barbara Pasquini are
gratefully acknowledged.
A special thank goes to Andreas Freund and Martin Mc Dermott,
for having provided us with their evolution code together with some
tuition of it, and for a critical reading of the manuscript.

\newpage

\centerline{\bf \Large Captions}
\hfill\break
{\bf Figure 1}: 
{The Proton Charge Form Factor, as given by 
Eq. (\protect{\ref{ffwf}}) in the IK model. }
\hfill\break
\hfill\break
{\bf Figure 2}:
{The GPD $H_q(x,\xi,\Delta^2)$, Eqs. 
(\protect{\ref{iku}}) and (\protect{\ref{ikd}}),
calculated for $\xi=0$
and three values of $\Delta^2$:
the full line corresponds to $\Delta^2=0$ GeV$^2$,
the dotted one to $\Delta^2=-0.2$ GeV$^2$ and the dashed one
to $\Delta^2=-0.5$ GeV$^2$. Left(right) panel: the $u$ ($d$) flavor 
distribution. 
Notice that, due to the chosen scale, the left and right panels should look
the same if the model under scrutiny were SU(6) symmetric.}
\hfill\break
\hfill\break
{\bf Figure 3}:
{The GPD $H(x,\xi,t)$, Eq. (21), calculated for
$\Delta^2=-0.2$ GeV$^2$ 
and three values of $\xi$:
the full line corresponds to $\xi=0$,
the dotted one to $\xi=0.1$ and the dashed one
to  $\xi=0.2$.
Left (right) panel: the $u$ ($d$) flavor distribution.
Notice that only the $x \geq \xi/2$ region is shown.
Due to scale used, the left and right panels should look
the same if the model under scrutiny were SU(6) symmetric.}
\hfill\break
\hfill\break
{\bf Figure 4}:
{The GPD $H(x,\xi,t)$, Eq. (21), calculated at fixed $\xi=0.1$ and 
$-\Delta^2$ in the range $-$0 GeV$^2$ -- 0.5 GeV$^2$ (left panel),
and at fixed $\Delta^2=-0.3$ GeV$^2$ and $\xi$ in the range 
0 -- 0.3 (right panel). 
}
\hfill\break
\hfill\break
{\bf Figure 5}:
{The GPD $H_d(x,0,\Delta^2)$, calculated within
the IK (full) and the h.o. (dashed) models, for
$\Delta^2 = 0$ GeV$^2$ (top), $-0.2$ GeV$^2$, $-0.5$ GeV$^2$ (bottom).
}
\hfill\break
\hfill\break
{\bf Figure 6}:
{The $NS$, $q$ flavor
GPD, $H^{NS}_q(x,\xi=0.1,\Delta^2=-0.2$ GeV$^2,Q^2=10$ GeV$^2)$ (full), 
evaluated evolving at NLO the initial IK model distribution
$H^{NS}_q(x,\xi=0,\Delta^2=-0.2$ GeV$^2$, $\mu_o^2=0.34$ GeV$^2)$ (dotted).
Left (right) panel: the $u$ ($d$) flavor distribution.
Only the $x \geq \xi/2$ region is shown.
Due to the chosen scale, the left and right panels should look
the same if the model under scrutiny were SU(6) symmetric.}
\hfill\break
\hfill\break
{\bf Figure 7}:
{As in Fig. 6, but at $\xi=0.2$, $\Delta^2=-0.5$ GeV$^2$.}
\hfill\break
\hfill\break
{\bf Figure 8}:
{The $NS$, $u$ flavor
GPD, $H^{NS}_u(x,\xi=0.1,\Delta^2=-0.2$ GeV$^2,Q^2)$,
with $Q^2$ in the range $\mu_o^2=0.34$ GeV$^2$ -- $Q^2$ = 10 GeV$^2$.
}

\newpage

\begin{figure}[h]
\vspace{16cm}
\includegraphics{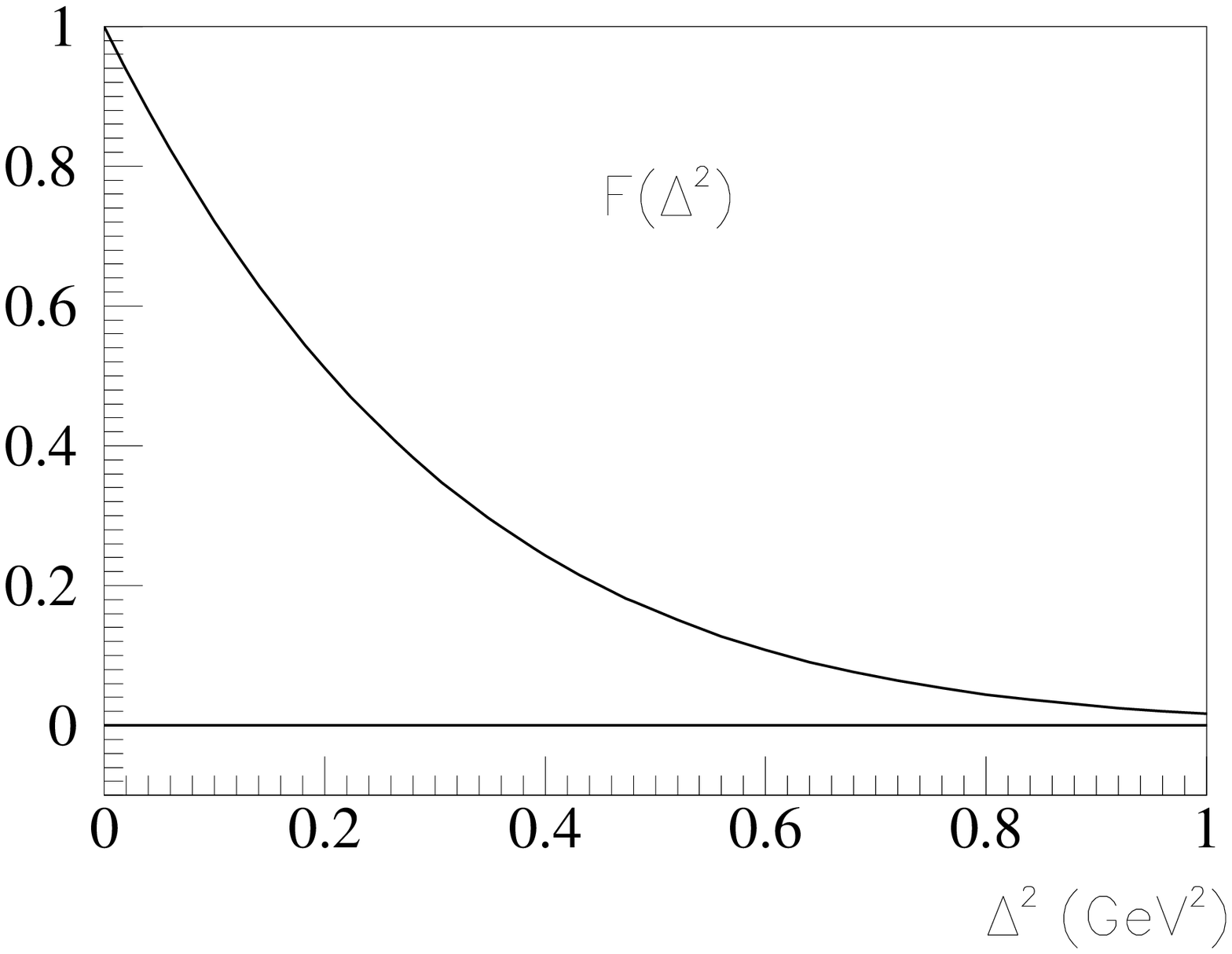}
\vskip -2.5cm
\caption{}
\end{figure}

\begin{figure}[htb]
\begin{minipage}[t] {77 mm}
\vspace{8.7cm}
\includegraphics{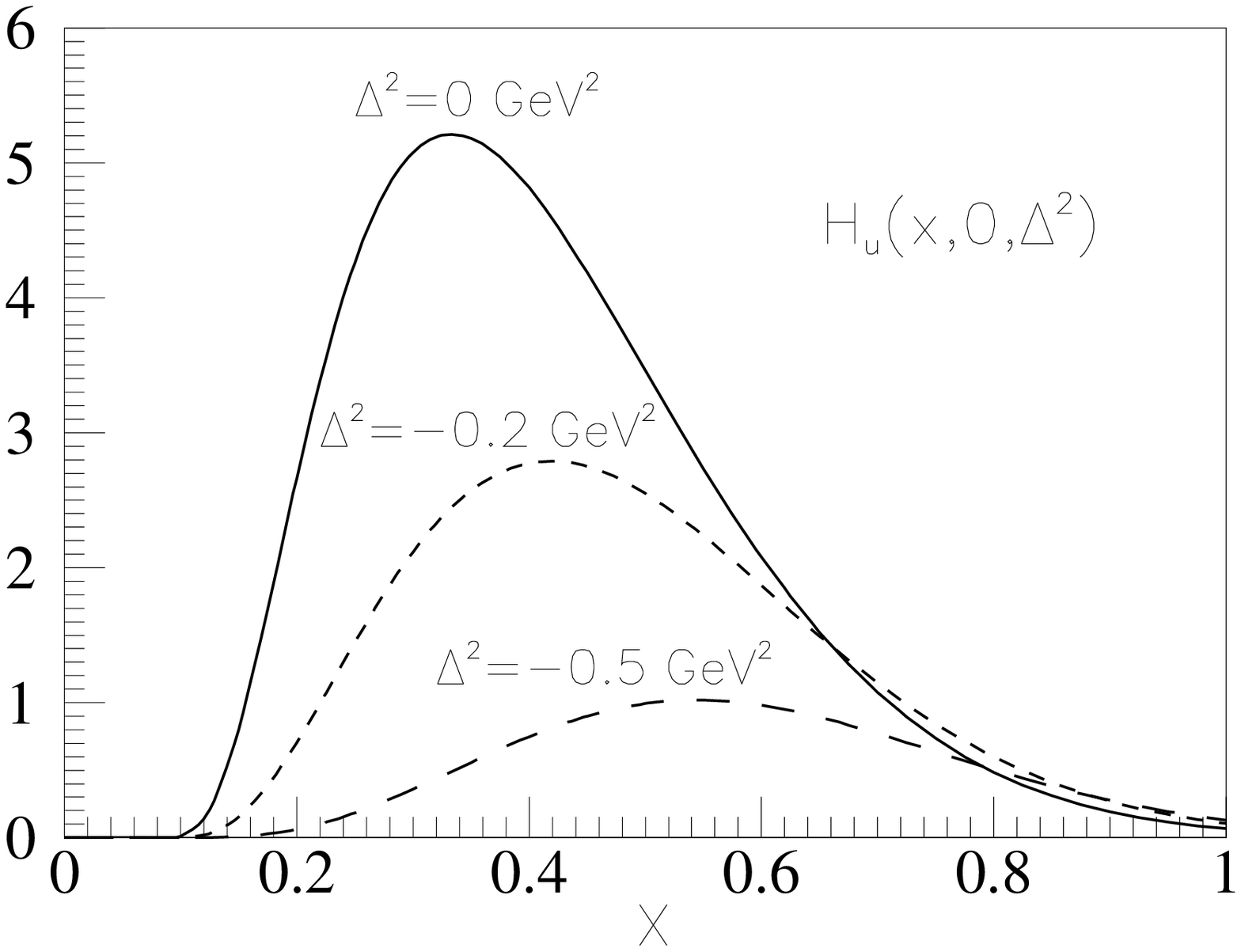}
\caption{ }
\end{minipage}
\hspace{\fill}
\begin{minipage}[t] {77 mm}
\vspace{8.7cm}
\includegraphics{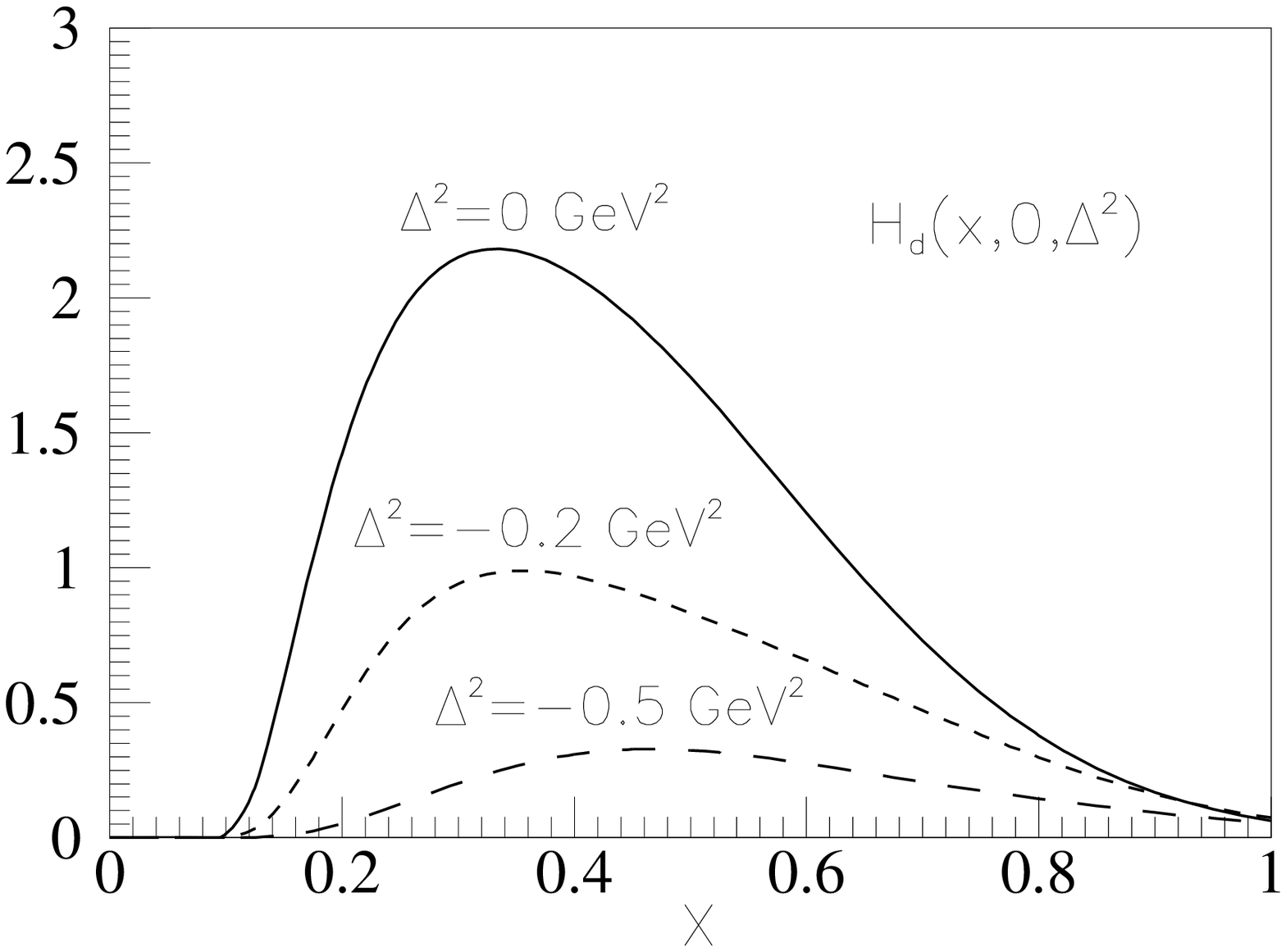}
\end{minipage}
\end{figure}

\begin{figure}[htb]
\begin{minipage}[t] {77 mm}
\vspace{8.7cm}
\includegraphics{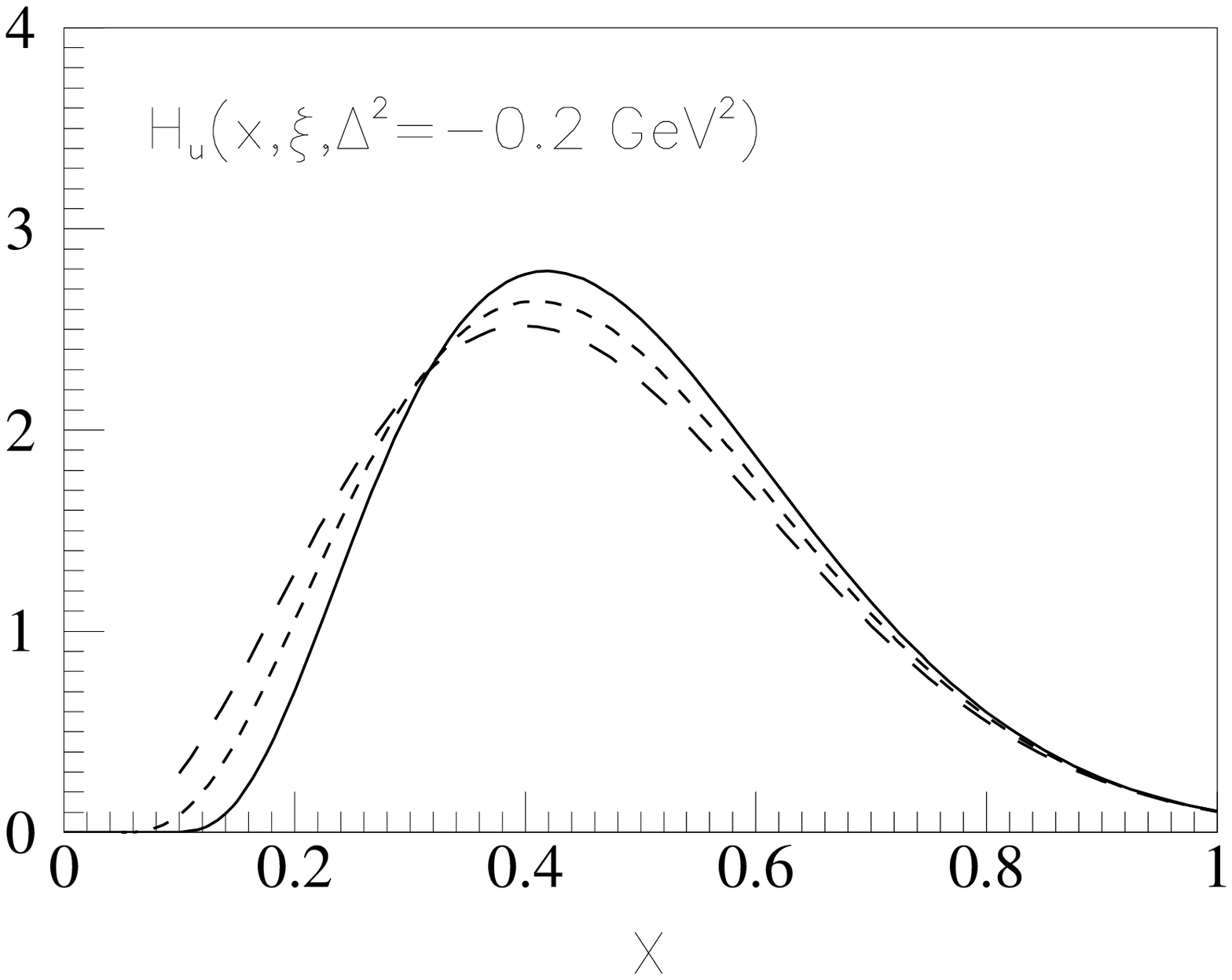}
\caption{ }
\end{minipage}
\hspace{\fill}
\begin{minipage}[t] {77 mm}
\vspace{8.7cm}
\includegraphics{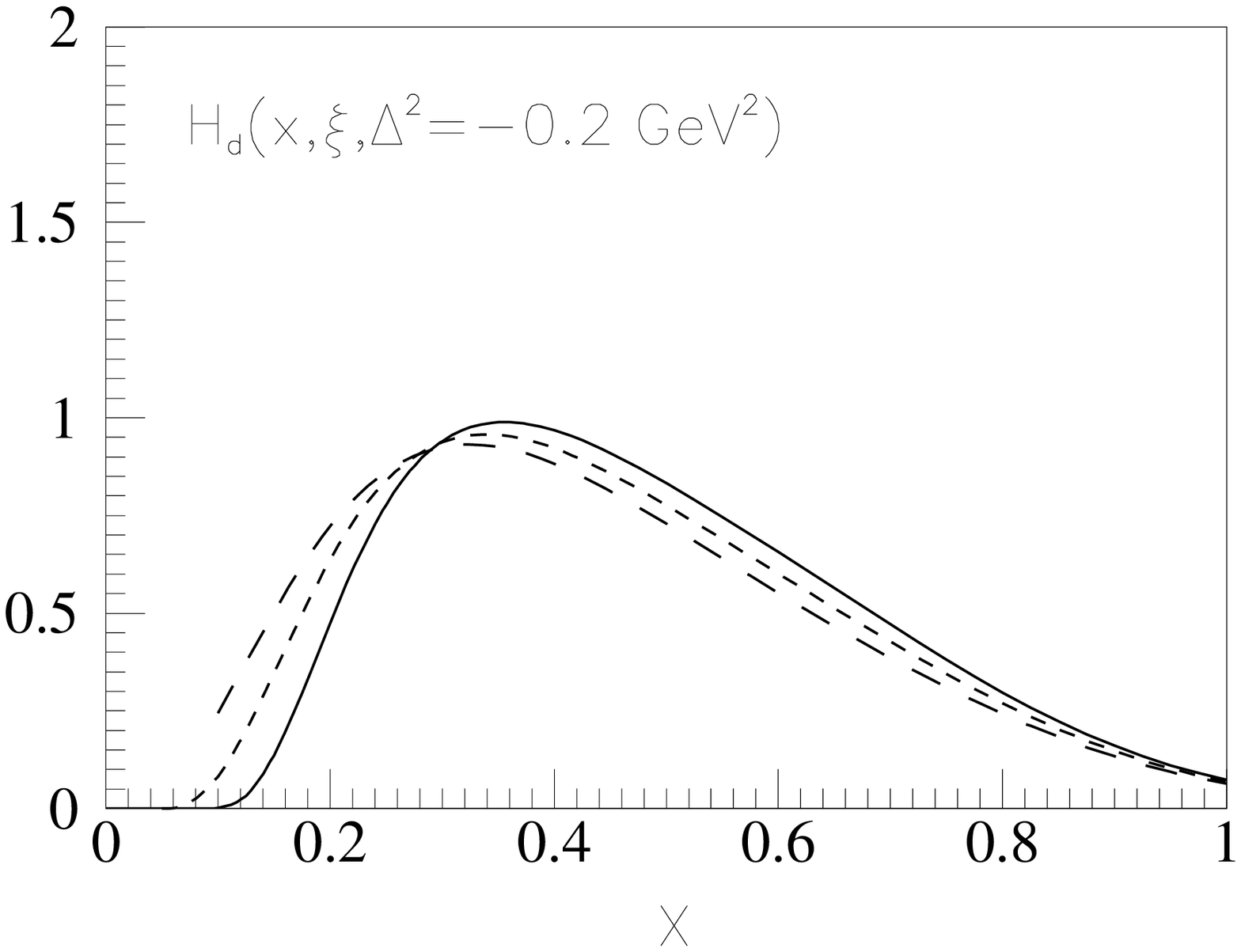}
\end{minipage}
\end{figure}

\newpage

\begin{figure}[htb]
\begin{minipage}[t] {77 mm}
\vspace{9.7cm}
\includegraphics{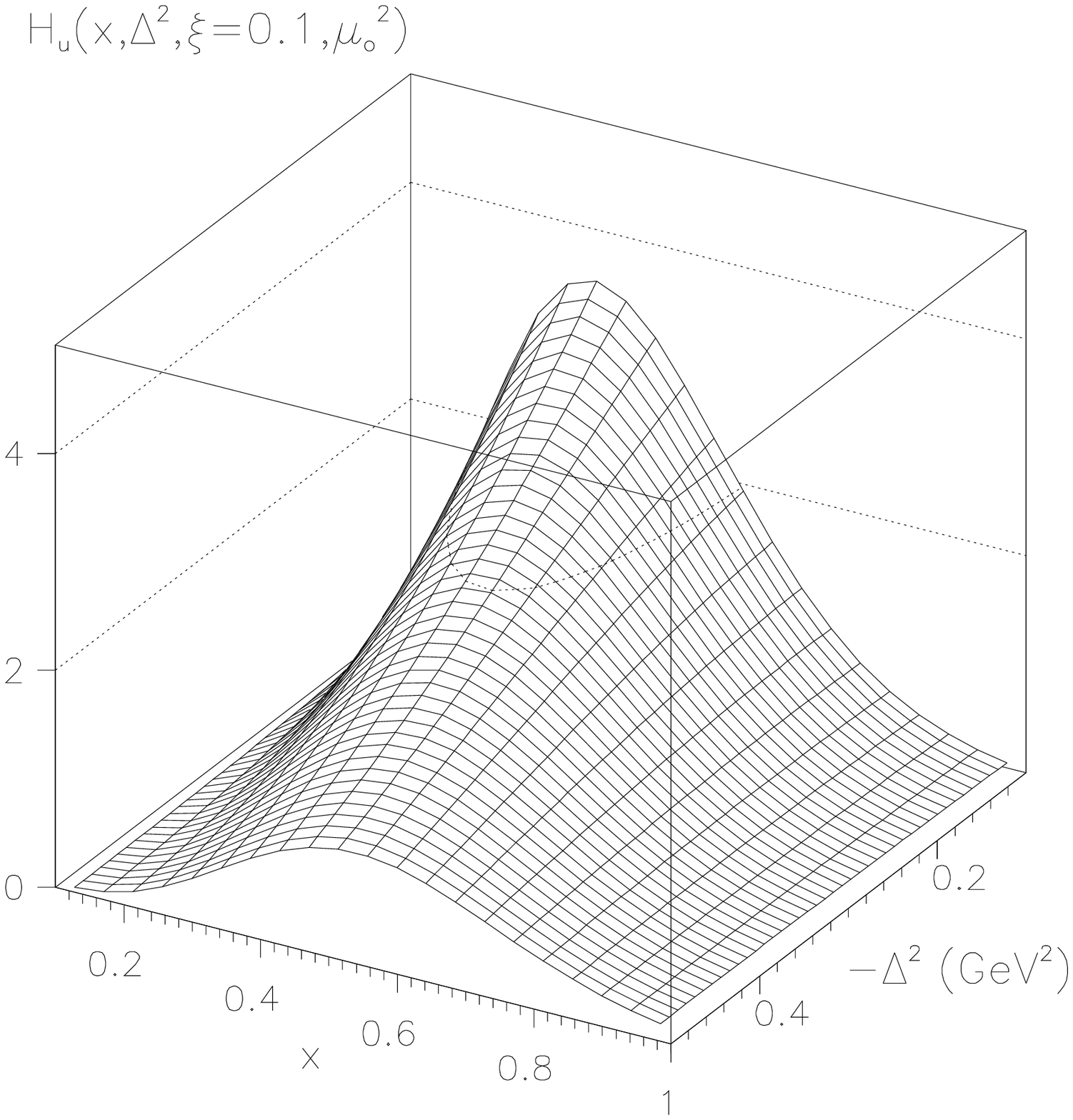}
\caption{ }
\end{minipage}
\hspace{\fill}
\begin{minipage}[t] {77 mm}
\vspace{9.7cm}
\includegraphics{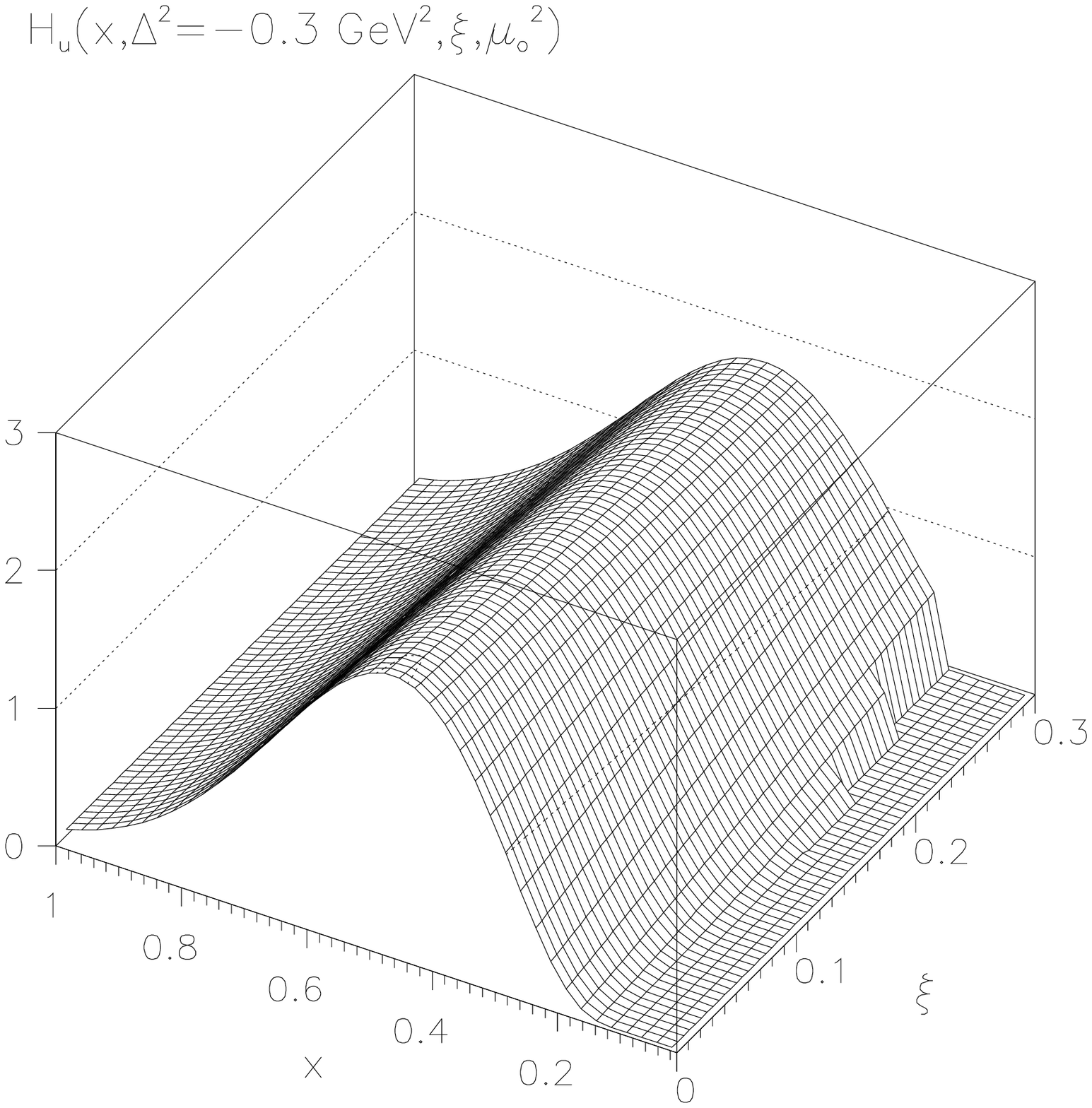}
\end{minipage}
\end{figure}

\begin{figure}[h]
\vspace{13cm}
\includegraphics{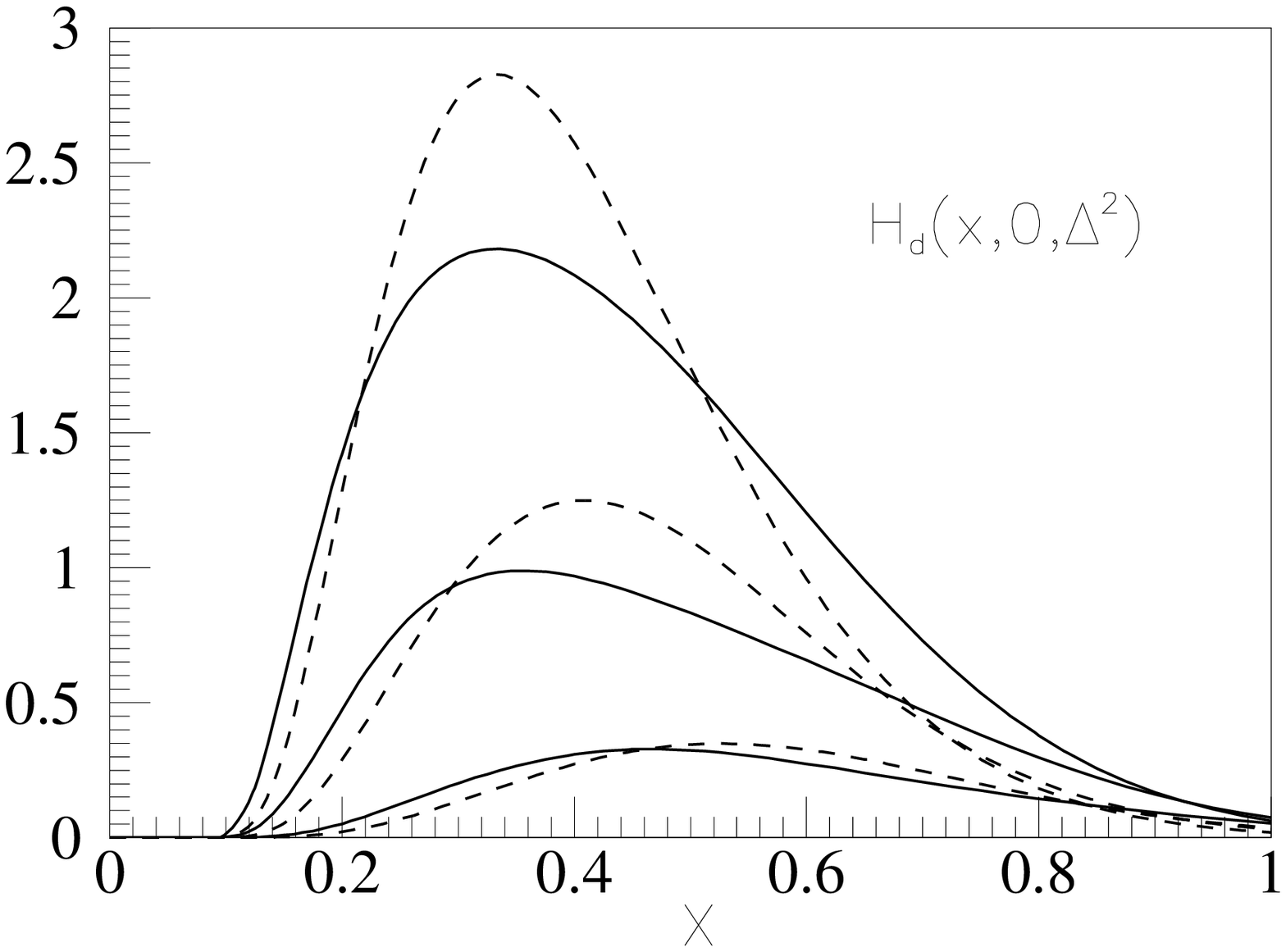}
\vskip -2.5cm
\caption{}
\end{figure}

\begin{figure}[htb]
\begin{minipage}[t] {77 mm}
\vspace{8.7cm}
\includegraphics{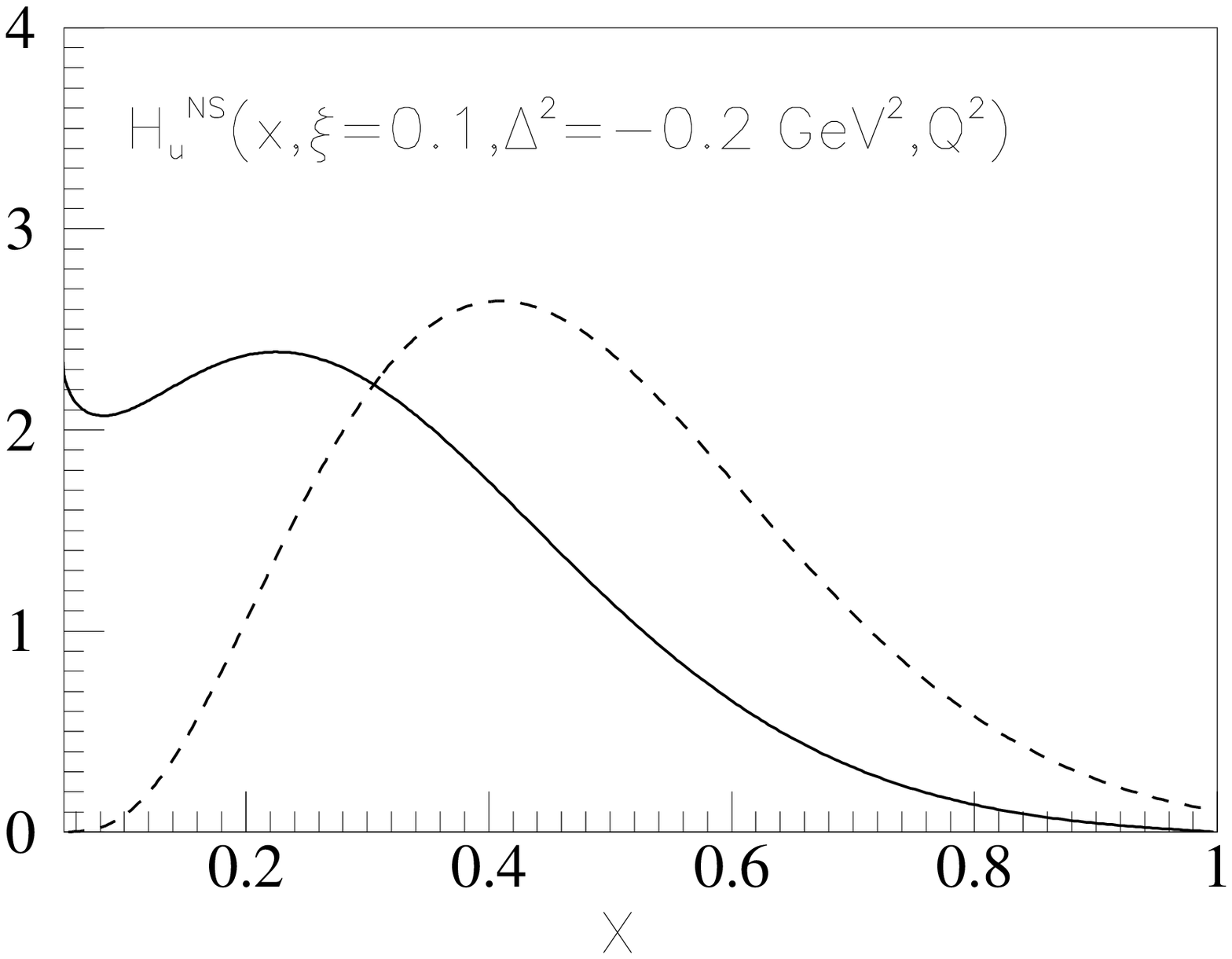}
\caption{ }
\end{minipage}
\hspace{\fill}
\begin{minipage}[t] {77 mm}
\vspace{8.7cm}
\includegraphics{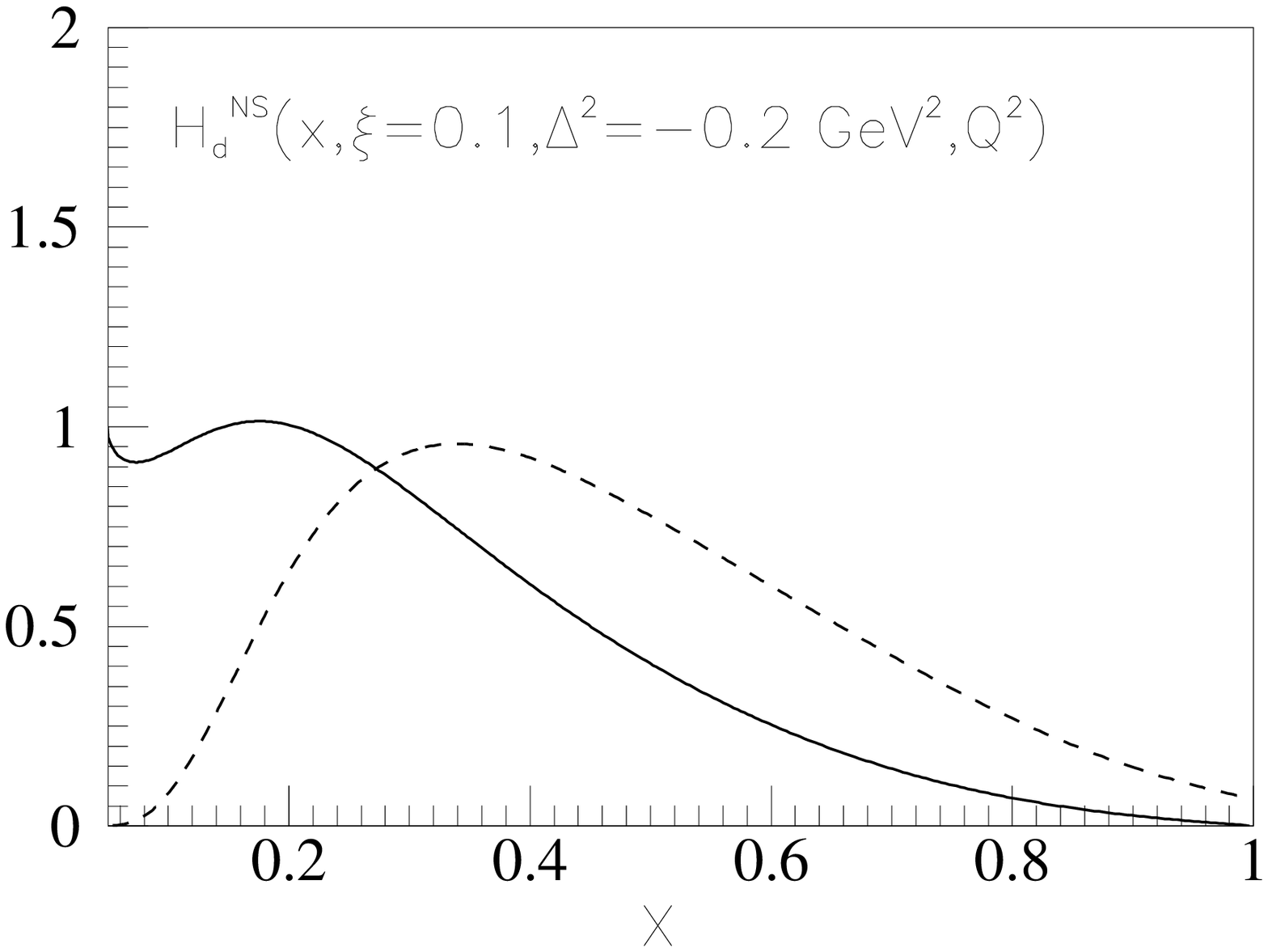}
\end{minipage}
\end{figure}

\begin{figure}[htb]
\begin{minipage}[t] {77 mm}
\vspace{8.7cm}
\includegraphics{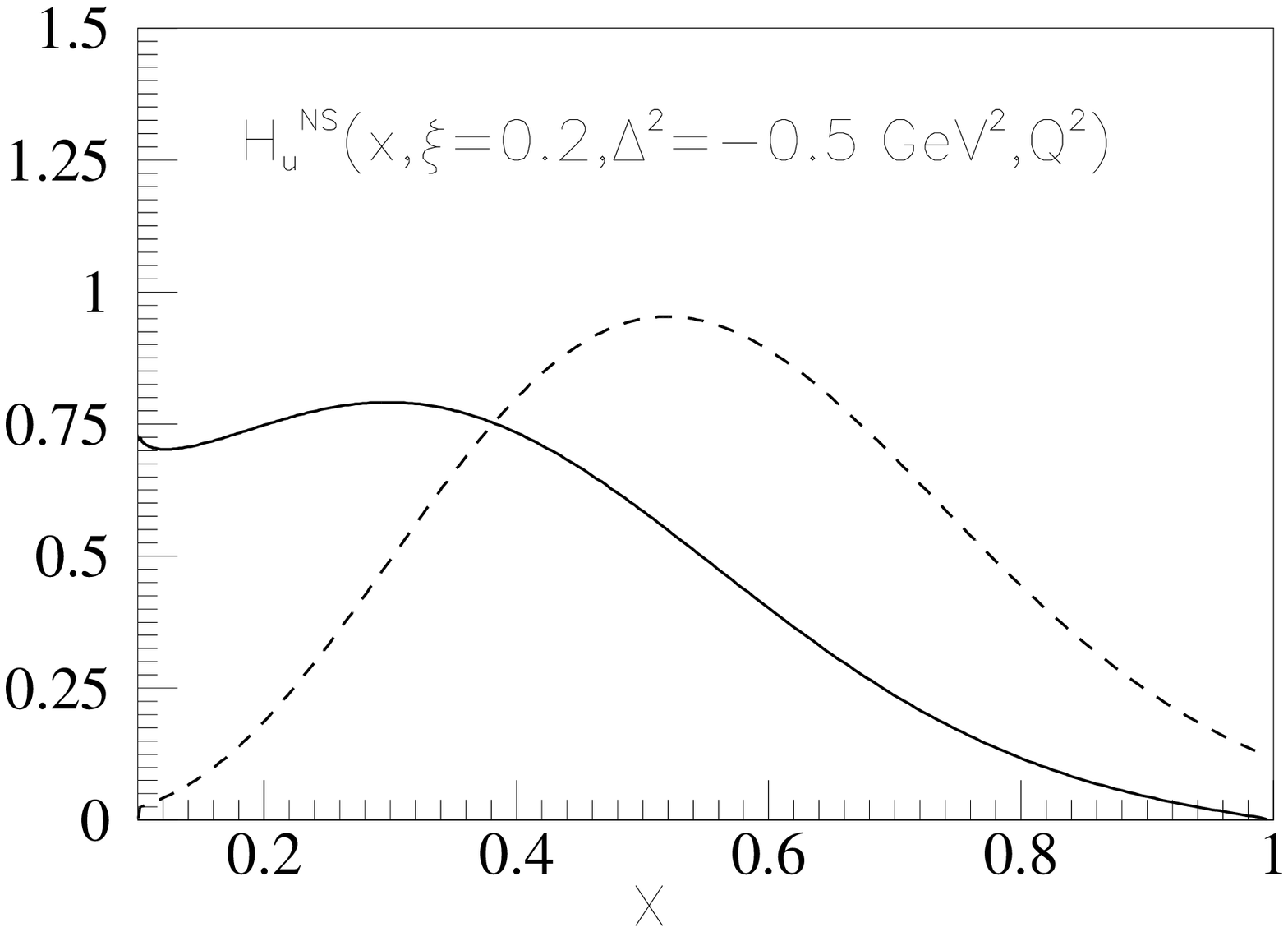}
\caption{ }
\end{minipage}
\hspace{\fill}
\begin{minipage}[t] {77 mm}
\vspace{8.7cm}
\includegraphics{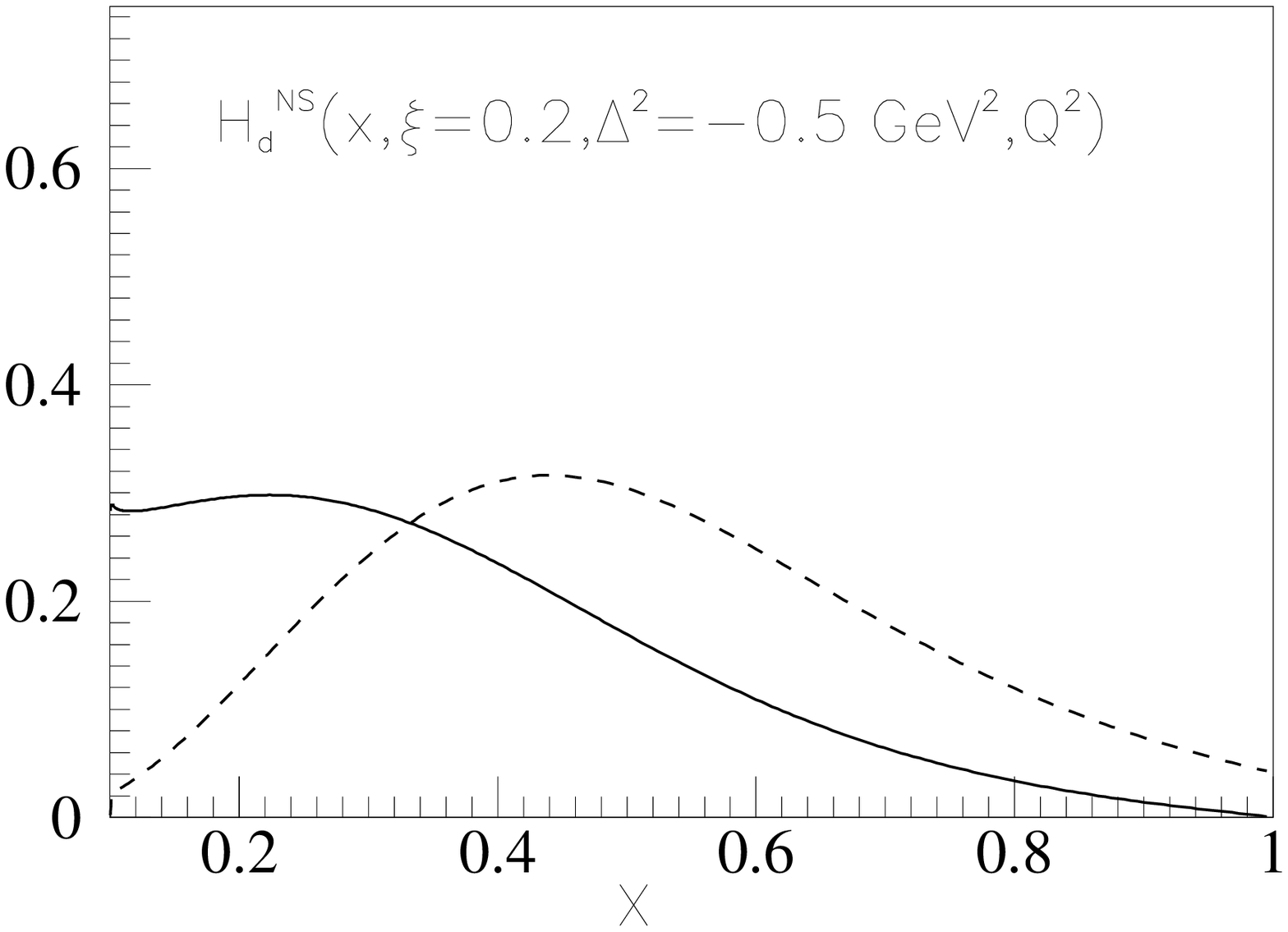}
\end{minipage}
\end{figure}

\begin{figure}[h]
\vspace{13cm}
\includegraphics{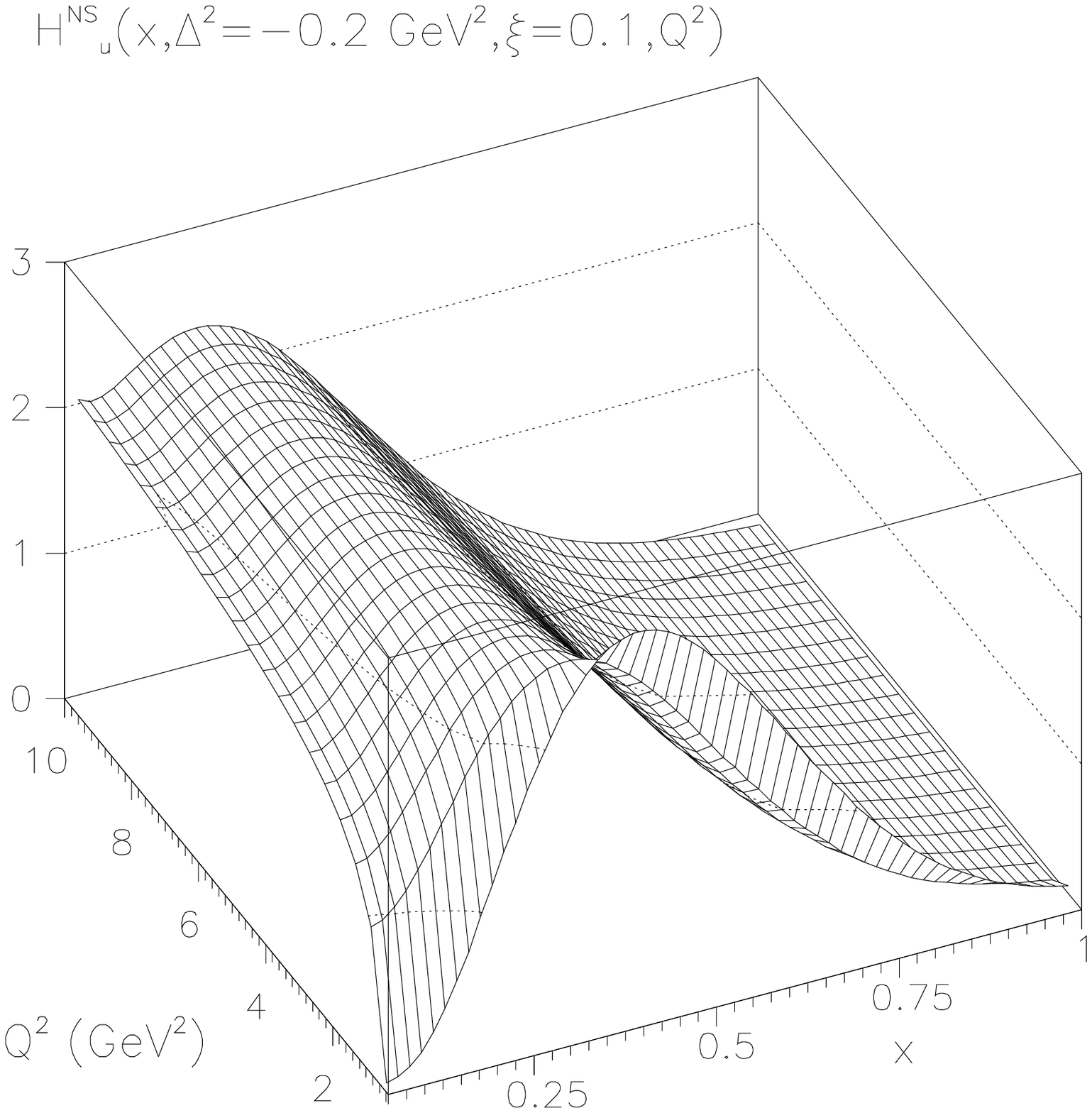}
\vskip 2.5cm
\caption{}
\end{figure}


\vskip 0.5cm


\begin{thebibliography}{99}
\bibitem{jig} X. Ji, J. Phys. G24 (1998) 1181. 
\bibitem{rag} A.V. Radyushkin, JLAB-THY-00-33,  
in M. Shifman, (ed.): At the frontier of particle physics, vol. 2, 1037-1099,
hep-ph/0101225.
\bibitem{pog} 
K. Goeke, M.V. Polyakov, M. Vanderhaeghen,
Prog. Part. Nucl. Phys.47 (2001) 401. 
\bibitem{burk1} M. Burkardt, Phys. Rev. D 62 (2000) 071503.
\bibitem{kroll}
By M. Diehl, T. Feldmann, R. Jakob, P. Kroll
Nucl. Phys. B596 (2001) 33; Phys. Lett. B460 (1999) 204. 
\bibitem{ji1} X. Ji, Phys. Rev. Lett. 78 (1997) 610.
\bibitem{jiprd} X. Ji, Phys. Rev. D 55 (1997) 7114.
\bibitem{jaffe} R.L. Jaffe, A.V. Manohar, Nucl. Phys. B 337 (1990) 509. 
\bibitem{ceb} ``The Science Driving the 12 GeV Upgrade of CEBAF'',
Jefferson Lab, February 2001.
\bibitem{esop} J.M. Laget, Nucl. Phys. A666, (2000) 336. 
\bibitem{meln} X. Ji, W. Melnitchouk, and X. Song, 
Phys. Rev. D 56 (1997) 5511.
\bibitem{vv} I.V. Anikin, D. Binosi, R. Medrano, S. Noguera,
and V. Vento, Eur.Phys.J.A14 (2002) 95. 
\bibitem{goe1} V.Yu. Petrov, P.V. Pobylitsa, M.V. Polyakov, 
I. Bornig, K. Goeke, C. Weiss, Phys. Rev. D57 (1998) 4325;
M. Penttinen, M.V. Polyakov, K. Goeke, Phys. Rev. D62 (2000) 014024. 
\bibitem{mill} B.C. Tiburzi, G.A. Miller 
Phys. Rev. C64 (2001) 065204; hep-ph/0109174.
\bibitem{freund} 
A. Freund, V. Guzey, Phys .Lett. B462 (1999) 178;
L. Frankfurt, A. Freund, V. Guzey, M. Strikman 
Phys. Lett. B418 (1998) 345.                             
\bibitem{rad1} I.V. Musatov, A.V. Radyushkin,
Phys. Rev. D61 (2000) 074027; A.V. Radyushkin 
Phys. Lett. B 449 (1999) 81. 
\bibitem{scha2}
A.V. Belitsky, B. Geyer, D. Muller, A. Sch\"afer
Phys. Lett. B421 (1998) 312;
A.V. Belitsky, D. Muller, L. Niedermeier, A. Sch\"afer
Phys. Lett. B437 (1998) 160; 
Nucl.Phys. B546 (1999) 279; 
Phys. Lett. B474 (2000) 163.
\bibitem{evo}
A.V. Belitsky, A. Freund, D. Muller Nucl. Phys. B574 (2000) 347; 
Phys. Lett. B493 (2000) 341; A. Freund, M.F. McDermott Phys. Rev. D65 (2002)
056012; Phys. Rev. D65 (2002) 091901; 
Phys. Rev. D65 (2002) 074008; Eur. Phys. J. C23 (2002) 651. 
\bibitem{scha1} 
A.V. Belitsky, A. Sch\"afer 
Nucl. Phys. B527 (1998) 235; 
N. Kivel, Maxim V. Polyakov, A. Sch\"afer, O.V. Teryaev 
Phys. Lett. B497 (2001) 73;
A.V. Belitsky, A. Kirchner, D. Muller, A. Sch\"afer 
Phys. Lett. B510 (2001) 117. 
\bibitem{burk2} M. Burkardt, hep-ph/0105324.
\bibitem{pape} G. Parisi, R. Petronzio, Phys. Lett. B 62 (1976) 331.
\bibitem{jaro} R.L. Jaffe, G.G. Ross, Phys. Lett. B 93 (1980) 313.
\bibitem{trvv} M. Traini, V. Vento, A. Mair and
A. Zambarda, Nucl. Phys. A 614 (1997) 472.
\bibitem{ik} N. Isgur and G. Karl, Phys. Rev. D 18 (1978) 4187, D 19 (1979)
 2653.
\bibitem{tbp} S. Scopetta and V. Vento, to be published.
\bibitem{dglap} V.N. Gribov and L. N. Lipatov, Sov. J. Nucl. Phys. 15, 78
(1972); L.N. Lipatov, Sov. J. Nucl. Phys. 20, 94 (1975);
G. Altarelli and G. Parisi, Nucl. Phys. B126 (1977) 298;
Yu.L. Dokshitser, Sov. Phys. JET, 46 (1977) 641.
\bibitem{erbl} A.V. Efremov and A.V. Radyushkin, Phys. Lett. B94
(1980) 245; S.J. Brodsky and G.P. Lepage, Phys. Rev. D22 (1980) 2157.
\bibitem{muld} P. Mulders, Phys. Rept. 185 (1990) 83. 
\bibitem{traini} L. Conci and M. Traini, Few-Body Systems 8,
123-136 (1990).
\bibitem{mmg} M.M. Giannini, Rep. Prog. Phys. 54 (1991) 453.
\bibitem{acmp} G. Altarelli, N. Cabibbo, L. Maiani and R. Petronzio,
Nucl. Phys. B 69 (1974) 531; G. Altarelli, S. Petrarca and
F. Rapuano, Phys. Lett. B 373 (1996) 200.
\bibitem{gianni} F. Cardarelli, E. Pace, G. Salme, S. Simula
Phys. Lett. B 357 (1995) 267.
\bibitem{scopetta} S. Scopetta, V. Vento and M. Traini,
Phys. Lett. B 421 (1998) 64; Phys. Lett. B 442 (1998) 28.
\bibitem{ik2} N. Isgur, G. Karl and R. Koniuk,
Phys. Rev. Lett. 41 (1978) 1269; N. Isgur, G. Karl and
J. Soffer, Phys. Rev. D 35 (1987) 1665.
\bibitem{tcm} M. Traini, L. Conci and U. Moschella,
Nucl. Phys. A544 (1992) 781.
\bibitem{web} http://durpdg.dur.ac.uk/hepdata/dvcs.html
\bibitem{g-bm} K.J. Golec-Biernat and A.D. Martin,
Phys. Rev. D59 (1999) 014029.
\bibitem{grv} M. Gl\"uck, E. Reya and A. Vogt, Eur. Phys. J. C5
(1998) 461 and reference therein.
 
\end{thebibliography}
\end{document}